\documentclass[journal, twoside, web]{ieeecolor}

\usepackage{lcsys, amsmath, amssymb, amsfonts, algorithmic, textcomp, bm, multirow, tikz, xcolor}
\usepackage{glossaries, cite, float, algorithm, booktabs, siunitx, threeparttable}
\usepackage[colorlinks=true, allcolors=black]{hyperref}
\usepackage[caption=false, font=footnotesize]{subfig}
\usepackage{mathtools}

\def\BibTeX{{\rm B\kern-.05em{\sc i\kern-.025em b}\kern-.08em
    T\kern-.1667em\lower.7ex\hbox{E}\kern-.125emX}}

\newacronym{MPC}{MPC}{Model Predictive Control}
\newacronym{ECI}{ECI}{Earth Centered Inertial}
\newacronym{LVLH}{LVLH}{Local Vertical Local Horizontal}
\newacronym{CW}{CW}{Clohessy-Wiltshire}
\newacronym{MIP}{MIP}{Mixed-Integer Program}
\newacronym{MILP}{MILP}{Mixed-Integer Linear Program}
\newacronym{OSAM}{OSAM}{On-Orbit Servicing, Assembly, and Manufacturing}
\newacronym{ADR}{ADR}{Active Debris Removal}

\begin{document}

\title{\vspace{-5mm}Convex MPC and Thrust Allocation with Deadband for Spacecraft Rendezvous}

\author{Pedro Taborda, Hugo Matias, Daniel Silvestre, Pedro Lourenço
\thanks{P. Taborda is with the Instituto Superior Técnico, University of Lisbon, Portugal (email: 
pedrotaborda04@tecnico.ulisboa.pt). }
\thanks{H. Matias is with Institute for Systems and Robotics, Instituto Superior Técnico, 
University of Lisbon (email: hugomatias@tecnico.ulisboa.pt).}
\thanks{D. Silvestre is with the School of Science and Technology from the NOVA University of 
Lisbon, with COPELABS from Lusófona University, and also with the Institute for Systems and 
Robotics, Instituto Superior Técnico, University of Lisbon 
(email: dsilvestre@isr.tecnico.ulisboa.pt).}
\thanks{P. Lourenço is with GMV, Alameda dos Oceanos no. 115, 1990-392 Lisbon, Portugal (email: palourenco@gmv.com).}
\thanks{This work was partially supported by the Fundação para a Ciência e a Tecnologia (FCT) 
through the Institute for Systems and Robotics (ISR), under the Laboratory for Robotics and 
Engineering Systems (LARSyS) project UIDB/50009/2020, through project PCIF/MPG/0156/2019 FirePuma, 
and through the COPELABS project UIDB/04111/2020.}}

\maketitle


\begin{abstract}
This paper delves into a rendezvous scenario involving a chaser and a target spacecraft, focusing 
on the application of Model Predictive Control (MPC) to design a controller capable of guiding the 
chaser toward the target. The operational principle of spacecraft thrusters, requiring a minimum 
activation time that leads to the existence of a control deadband, introduces mixed-integer 
constraints into the optimization, posing a considerable computational challenge due to the 
exponential complexity on the number of integer constraints. We address this complexity by 
presenting two solver algorithms that efficiently approximate the optimal solution in significantly 
less time than standard solvers, making them well-suited for real-time applications.
\end{abstract}


\section{Introduction} \label{Section:Introduction}

\IEEEPARstart{S}{pacecraft} rendezvous is an essential task in space exploration, required for 
many missions, including docking, \gls{ADR} \cite{bonnal2013active}, and \gls{OSAM} operations 
\cite{arney2021orbit}. Given the safety standards and high cost of these missions, the space 
industry tends to be very conservative in nature, favoring the use of outdated tools and 
theoretical methods with a significant focus on safety, robustness, and trust over performance. 
Solutions often rely on open-loop control, where errors accumulate and are only rectified during
planned correction maneuvers \cite{fehse2003automated}. The thrust allocation is also typically
handled by a distinct unit that determines firing intervals for the thrusters 
\cite{bezerra2021optimal, ankersen2005optimization}.

In this context, \gls{MPC} has gained considerable traction in recent decades 
\cite{hartley2015tutorial, weiss2015model, silvestre2023model, zhu2018robust, leomanni2021sum}. 
\gls{MPC} provides continuous error correction, which is a critical feature for scenarios 
involving repetitive rendezvous phases, such as in-orbit assembly of large structures or servicing 
missions \cite{brizguidance}. In addition, \gls{MPC} can implicitly handle the allocation problem 
and minimize fuel consumption. This is particularly significant as the far-range stages of the 
rendezvous process substantially contribute to delta-V expenditure.

Nevertheless, the computational complexity of \gls{MPC} can be a significant challenge in the space 
flight context, as thrusters usually require a minimum firing duration, which introduces 
mixed-integer constraints into the optimization \cite{ankersen2005optimization}. Additionally, the 
onboard computational power, generally limited by the available power budget and the necessary 
hardening against space radiation, also restricts the application of \gls{MPC} 
\cite{lovelly:2018:BenchmarkingAnalysisSpaceGrade}. While some research has been successful in 
reducing the computational burden of \gls{MPC} for some space-flight applications 
\cite{accikmecse2011lossless, malyuta2020lossless}, there is still a gap in addressing this 
challenge for the particular case of spacecraft rendezvous. 

A common and straightforward approach for solving \gls{MPC} problems involving discrete actuation 
states (e.g., on or off) is to use binary variables for the actuation states and employ a \gls{MIP} 
solver. This is the most direct and standard approach, employed in several applications such as 
\cite{hespanhol2019structure}. An \gls{MIP} is an optimization problem involving discrete and 
continuous variables that can be solved using algorithms like the branch-and-bound algorithm, 
employed by the widely-used Gurobi solver \cite{gurobi}, and its variants for \gls{MPC} 
applications \cite{bemporad2018numerically}. However, these algorithms are not suitable for 
real-time applications as they present worst-case exponential complexity, which arises from 
the nature of solving an \gls{MIP}, requiring extensive search to identify a feasible solution.

This paper introduces an \gls{MPC} strategy for spacecraft rendezvous, focusing on orbital control 
rather than attitude control and docking, which is addressed in more detail in \cite{Yang2018}. We 
address the computational challenges arising from mixed-integer constraints introduced by 
spacecraft actuators (e.g., thruster deadband due to minimum impulse bit 
\cite[Chapter 3]{brown:1996:SpacecraftPropulsion}) by proposing two solver algorithms that 
efficiently approximate the optimal solution, significantly reducing computation time compared to 
standard solvers. 

The remaining sections are organized as follows. We define the control problem and describe the 
spacecraft dynamics in Section \ref{Section:ProblemFormulation}. Section 
\ref{Section:ProposedSolution} describes the \gls{MPC} approach and the proposed solver algorithms, 
and Section \ref{Section:SimulationResults} presents simulation results. Finally, Section 
\ref{Section:Conclusion} summarizes conclusions.

\subsubsection*{Notation}

We designate the set of integers from $i$ to $j$ ($i \leq j$) as $\mathbb{Z}_{[i, j]}$. The set
$\mathcal{S}^M$ denotes the $M$-ary Cartesian power of a set $\mathcal{S}$. Moreover, 
$\mathbf{0}_{m \times n}$ and $\mathbf{1}_{m \times n}$ denote, respectively, the matrices of zeros 
and ones with dimension $m \times n$ (if $n=1$, $n$ is omitted), and $\mathbf{I}_n$ is the identity 
matrix of size $n$. The set of positive-definite matrices with size $n$ is designated as 
$\mathbb{R}^{n \times n}_{\succ 0}$.


\section{Preliminaries \& Problem Statement} \label{Section:ProblemFormulation}

\begin{figure}[H] 
    \centering 
    \includegraphics[width=\linewidth]{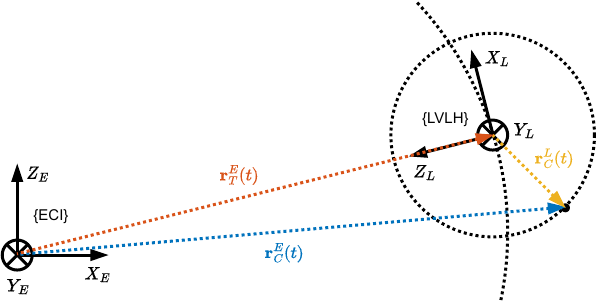} 
    \caption{\gls{ECI} and \gls{LVLH} reference frames.} 
    \label{Figure:ReferenceFrames}  
\end{figure}

We aim to develop an \gls{MPC} controller for a chaser spacecraft attempting to rendezvous, 
using chemical propulsion, with a target spacecraft that is in a circular orbit around the 
Earth. We assume the gravitational pull caused by the spacecraft to be negligible. The motion of 
the spacecraft is described in the \gls{ECI} and \gls{LVLH} frames depicted in Fig. 
\ref{Figure:ReferenceFrames}. The \gls{ECI} frame has its origin at the Earth's center and is 
nonrotating. Conversely, the \gls{LVLH} frame has its origin at the target's center, and its axes 
are defined as follows: the $z$-axis points from the target to the Earth's center, the $x$-axis has 
the target's velocity direction, and the $y$-axis is perpendicular to the orbital plane.

\subsection{Full Dynamics Model}

Since the goal is controlling the chaser spacecraft's relative position to the target, it is useful 
to describe the motion of the spacecraft in the \gls{LVLH} frame, as the target remains stationary 
in this frame. At each time instant $t$, the chaser's position in the \gls{LVLH} frame, 
$\mathbf{r}_C^L(t) \in \mathbb{R}^3$, is given by
\begin{equation}
    \mathbf{r}_C^L(t) = \mathbf{R}^\top(t)(\mathbf{r}_C^E(t) - \mathbf{r}_T^E(t)),
    \label{Equation:ChaserPosition}
\end{equation}
where $\mathbf{r}_C^E(t)$ and $\mathbf{r}_T^E(t)$ denote the positions of the chaser and target 
spacecraft in the \gls{ECI} frame, and $\mathbf{R}(t) \in \text{SO}(3)$ is the rotation matrix 
from the \gls{LVLH} frame to the \gls{ECI} frame. Taking now the second-order derivative of 
\eqref{Equation:ChaserPosition} with respect to time, it can be shown that the chaser dynamics in 
the \gls{LVLH} frame are described by
\begin{equation}
    \begin{aligned}
        \Ddot{\mathbf{r}}_C^L(t) = &-2\bm{\omega}\times\Dot{\mathbf{r}}_C^L(t) 
        - \bm{\omega}\times(\bm{\omega}\times\mathbf{r}_C^L(t))\\
        &+ \mathbf{R}^\top(t)(\Ddot{\mathbf{r}}_C^E(t) - \Ddot{\mathbf{r}}_T^E(t)),
    \end{aligned}
    \label{Equation:ChaserAccelerationLVLH}
\end{equation}
where $\bm{\omega} \in \mathbb{R}^3$ is the angular velocity of the \gls{LVLH} frame with respect 
to the \gls{ECI} frame. Also, by Newton's Second Law and Law of Universal Gravitation, 
$\Ddot{\mathbf{r}}_C^E(t)$ and $\Ddot{\mathbf{r}}_C^E(t)$ are given by
\begin{equation}
    \begin{aligned}
        \Ddot{\mathbf{r}}_C^E(t) &= -\frac{Gm_E}{\|\mathbf{r}_C^E(t)\|^3}\mathbf{r}_C^E(t) 
        + \frac{1}{m_C}\mathbf{R}(t)\mathbf{u}(t),\\
        \Ddot{\mathbf{r}}_T^E(t) &= -\frac{Gm_E}{\|\mathbf{r}_T^E(t)\|^3}\mathbf{r}_T^E(t),
    \end{aligned}
    \label{Equation:AccelerationsECI}
\end{equation}
where $G$ is the gravitational constant, $m_E$ is the Earth's mass, $m_C$ is the chaser's 
mass, and $\mathbf{u}(t) \in \mathbb{R}^3$ denotes the forces acting on the chaser spacecraft due 
to the actuation, expressed in the \gls{LVLH} frame. In addition, we highlight that 
$\mathbf{r}_T^E(t)$, $\mathbf{R}(t)$ and $\bm{\omega}$ can be specified \textit{a priori} based on 
the target's circular orbit, which is assumed to be known.

\subsection{Simplified Model}

Given the highly nonlinear nature of the model described by \eqref{Equation:ChaserAccelerationLVLH} 
and \eqref{Equation:AccelerationsECI}, it is beneficial for the \gls{MPC} controller to consider a 
simplified model for enhanced efficiency. By noting that the distance between the two spacecraft is 
much smaller than the orbital radius of the target and then linearizing the full relative dynamics 
in the \gls{LVLH} frame around the circular orbit, the chaser's motion can be approximately 
described by the well-known \gls{CW} equations \cite[Appendix A]{fehse2003automated}. These can be 
expressed in linear state-space form as
\begin{equation}
    \Dot{\Tilde{\mathbf{x}}}(t) = \mathbf{A}\Tilde{\mathbf{x}}(t) + \mathbf{B}\mathbf{u}(t),
    \label{Equation:CW}
\end{equation}
where $\mathbf{A}$ and $\mathbf{B}$ are given by
\begin{equation}
    \mathbf{A} = 
    \begin{bmatrix}
        0 & 0 & 0 & 1 & 0 & 0\\
        0 & 0 & 0 & 0 & 1 & 0\\
        0 & 0 & 0 & 1 & 0 & 1\\
        0 & 0 & 0 & 0 & 0 & 2\omega\\
        0 & -\omega^2 & 0 & 0 & 0 & 0\\
        0 & 0 & 3\omega^2 & -2\omega & 0 & 0
    \end{bmatrix},\,\,
    \mathbf{B} = \frac{1}{m_C}
    \begin{bmatrix}
        \mathbf{0}_{3\times3}\\
        \mathbf{I}_{3\times3}
    \end{bmatrix},
\end{equation}
and with state $\Tilde{\mathbf{x}}(t) = [\Tilde{\mathbf{r}}_C^L(t)^\top\,\, 
\Dot{\Tilde{\mathbf{r}}}_C^L(t)^\top]^\top$, where tilde notation distinguishes the 
state of the approximate model from the state
$\mathbf{x}(t) = [\mathbf{r}_C^L(t)^\top\,\, \Dot{\mathbf{r}}_C^L(t)^\top]^\top$ of the full 
dynamics model.

\subsection{Actuation and Model Discretization}

Spacecraft propulsion systems vary in complexity and capabilities. Such systems encompass chemical 
and electric propulsion, achieving high and low thrust, respectively 
\cite{brown:1996:SpacecraftPropulsion}. Within chemical propulsion, necessary for rendezvous, 
thrusters operate on a shared principle: a propellant is heated and expelled through a nozzle, 
generating a fixed and constant thrust. In addition, thrusters have a minimum operational time that 
they should be active for (minimum impulse bit \cite{brown:1996:SpacecraftPropulsion}).

At each sampling step $k$, each thruster can either generate a single force pulse from the 
beginning of the sampling interval or remain inactive. Therefore, the control signal produced by a 
single thruster $i$, $\mathbf{u}_i(t)$, is piecewise constant and given by
\begin{equation}
    \mathbf{u}_i(t) = 
    \begin{cases}
        \mathbf{f}_i,\,\, \text{if}\,\, t \in [kh, kh + s_i[k]),\\
        \mathbf{0}_3,\,\, \text{if}\,\, t \in [kh + s_i[k], (k+1)h),
    \end{cases}
\end{equation}
for all $k \in \mathbb{Z}_0^+$, where $\mathbf{f}_i \in \mathbb{R}^3\setminus\{\mathbf{0}_3\}$ 
denotes the constant thrust force generated by the thruster, $h$ denotes the sampling period, and 
$s_i[k] \in \{0\} \cup [h_\text{min}, h]$ is the pulse duration, which must have a minimum value of 
$h_\text{min} \in (0, h]$ when the thruster is activated. Hence, by noting that the overall control 
signal is given by $\mathbf{u}(t) = \sum_{i=1}^M \mathbf{u}_i(t)$, the linear continuous-time model 
in \eqref{Equation:CW} can be exactly discretized as
\begin{equation}
    \begin{aligned}[b]
        \Tilde{\mathbf{x}}[k+1] 
        &= e^{\mathbf{A}h}\Tilde{\mathbf{x}}[k] + \sum_{i=1}^{M}\int_{kh}^{(k+1)h}
        e^{\mathbf{A}((k+1)h-\tau)}\mathbf{B}\mathbf{u}_i(\tau)\,d\tau\\
       &= \mathbf{\Phi}\Tilde{\mathbf{x}}[k] + \mathbf{\Phi}\sum_{i=1}^{M}\int_0^{s_i[k]}
       e^{-\mathbf{A}\tau}\,d\tau\mathbf{B}\mathbf{f}_i\\
        &= \mathbf{\Phi}\Tilde{\mathbf{x}}[k] + 
        \mathbf{\Phi}\sum_{i=1}^{M}\mathbf{G}(s_i[k])\mathbf{B}\mathbf{f}_i,
    \end{aligned}
    \label{Equation:CWDiscrete}
\end{equation}
with $\mathbf{\Phi} := e^{\mathbf{A}h}$ and $\mathbf{G}(s) := \int_0^{s}e^{-\mathbf{A}\tau}\,d\tau$,
$\forall s \in \mathbb{R}$.

\subsection{Linearization of the Actuator Dynamics}

For the discrete-time model in \eqref{Equation:CWDiscrete}, the state transition matrix 
$e^{\mathbf{A}h}$ is given by
\begin{equation}
    e^{\mathbf{A}h} = 
    \scalebox{0.55}{$
    \begin{bmatrix}
        1 & 0 & 6(\omega h- \sin(\omega h)) & \frac{4}{\omega}\sin(\omega h)-3h & 0 & 
        \frac{2}{\omega}(1-\cos(\omega h))\\
        0 & \cos(\omega h) & 0 & 0 & \frac{1}{\omega}\sin(\omega h) & 0 \\
        0 & 0 & 4 -3\cos(\omega h) & \frac{2}{\omega}(\cos(\omega h) - 1) & 0 
        & \frac{1}{\omega}\sin(\omega h) \\
        0 & 0 & 6\omega(1-\cos(\omega h)) & 4\cos(\omega h)-3 & 0 & 2\sin(\omega h) \\
        0 & -\omega\sin(\omega h) & 0 & 0 & \cos(\omega h) & 0 \\
        0 & 0 & 3\omega\sin(\omega h) & -2\sin(\omega h) & 0 & \cos(\omega h)
    \end{bmatrix}$}.
\end{equation}
As a result, $\mathbf{G}$ is a nonlinear function, rendering the model in 
\eqref{Equation:CWDiscrete} nonlinear with respect to the control variables $s_i[k]$. As we aim to 
build an \gls{MPC} constrained by these dynamics at each time step, it is beneficial to linearize 
$\mathbf{G}$ so that linear solvers can be used to solve the optimization problem. Consequently, 
linearizing $\mathbf{G}$ around a point $s_0$ leads to
\begin{equation}
    \begin{aligned}
        \mathbf{G}(s) \simeq \mathbf{G}(s_0) + e^{-\mathbf{A}s_0}(s - s_0),
    \end{aligned}
\end{equation}
resulting in the affine discrete-time model
\begin{equation}
        \Tilde{\mathbf{x}}[k+1] = \mathbf{\Phi}\Tilde{\mathbf{x}}[k] 
        + \mathbf{\Gamma}\mathbf{s}[k] + \mathbf{d},
    \label{Equation:LinearDiscrete}
\end{equation}
where $\mathbf{s}[k] := [s_1[k] \dots s_M[k]]^\top$, 
$\mathbf{\Gamma} = e^{\mathbf{A}(h-s_0)}[\mathbf{B}\mathbf{f}_1 \dots 
\mathbf{B}\mathbf{f}_M]$ and 
$\mathbf{d} = e^{\mathbf{A}h}(\mathbf{G}(s_0) - e^{-\mathbf{A}h}s_0) 
\mathbf{\Gamma}\mathbf{1}_{M}$.


\section{\texorpdfstring{Model Predictive Control \\with Mixed-Integer Constraints}{Model 
Predictive Control with Mixed-Integer Constraints}} \label{Section:ProposedSolution}

Considering the setup described in the preceding section, at every discrete-time instant $k$, for a 
given initial state $\mathbf{x}[k]$ of the chaser spacecraft, the \gls{MPC} controller relies on 
the following optimization problem for a given horizon $N$:
\begin{equation} 
    \begin{aligned}
        \underset{\hat{\mathbf{x}}[\cdot], \hat{\mathbf{s}}[\cdot]}{\text{minimize}} 
        \quad & J(\hat{\mathbf{x}}[\cdot], \hat{\mathbf{s}}[\cdot])\\
        \text{subject to} \quad & \hat{\mathbf{x}}[0] = \mathbf{x}[k],\\
        & \hat{\mathbf{x}}[n+1] = 
        \mathbf{\Phi}\hat{\mathbf{x}}[n] + \mathbf{\Gamma}\hat{\mathbf{s}}[n] + \mathbf{d},\, 
        \forall n \in \mathbb{Z}_{[0, N-1]},\\
        & \hat{\mathbf{s}}[n] \in (\{0\} \cup [h_\text{min}, h])^M,\,\,
        \forall n \in \mathbb{Z}_{[0, N-1]},
    \end{aligned}
    \label{Equation:MPC}
\end{equation}
where $\hat{\mathbf{x}}[\cdot]$ and $\hat{\mathbf{s}}[\cdot]$ are the optimization variables 
matching the predicted state and control sequences over the horizon, and the cost function $J$ 
should translate the mission objectives. For the problem at hand, a possible definition of the cost 
function is
\begin{equation}
    J(\hat{\mathbf{x}}[\cdot], \hat{\mathbf{s}}[\cdot]) 
    = \hat{\mathbf{x}}[N]^\top\mathbf{Q}\hat{\mathbf{x}}[N] 
    + \sum_{n = 0}^{N-1}\mathbf{1}_M^\top\hat{\mathbf{s}}[n],
    \label{Equation:CostFunction}
\end{equation}
where $\mathbf{Q} \in \mathbb{R}^{6\times6}_{\succ 0}$. The first component of 
$\eqref{Equation:CostFunction}$ expresses the objective of reaching the target spacecraft, located
at the origin of the \gls{LVLH} frame. The second penalizes the control effort, obtained by summing 
the activation times of all thrusters over the prediction horizon.

Owing to the constraints on each actuator's activation time $\hat{s}_i[n]$, the optimization 
problem \eqref{Equation:MPC} becomes a \gls{MILP}. The most straightforward approach to address 
this problem involves using a \textbf{Standard} \gls{MILP} solver, such as Gurobi, which guarantees 
the optimal solution. However, employing \textbf{Standard} \gls{MILP} solvers becomes impractical 
for real-time applications as they present worst-case exponential complexity on the number of 
integer constraints ($N \times M$). In this letter, we propose two alternative, more 
computationally efficient methods capable of producing solutions similar to the optimal one.

\subsection{Projected Algorithm}

The first proposed solver algorithm is denoted as \textbf{Projected}. Consider the auxiliary 
variables $\bm{\alpha}, \bm{\beta} \in \{0, 1\}^{M}$ such that, if $\alpha_i = 1$, then 
$\hat{s}_i[0] \in [h_\textrm{min}, h]$, and if $\beta_i = 1$, then $\hat{s}_i[0] = 0$. Also, 
$\alpha_i$ and $\beta_i$ are such that $\alpha_i\beta_i = 0,\,\, \forall i \in 
\mathbb{Z}_{[1, M]}$. The algorithm consists of sequentially solving the relaxed problem
\begin{equation} 
    \begin{aligned}
        \underset{\hat{\mathbf{x}}[\cdot], \hat{\mathbf{s}}[\cdot]}{\text{minimize}} 
        \quad & J(\hat{\mathbf{x}}[\cdot], \hat{\mathbf{s}}[\cdot])\\
        \text{subject to} \quad & \hat{\mathbf{x}}[0] = \mathbf{x}[k],\\
        & \hat{\mathbf{x}}[n+1] = 
        \mathbf{\Phi}\hat{\mathbf{x}}[n] + \mathbf{\Gamma}\hat{\mathbf{s}}[n] + \mathbf{d},\, 
        \forall n \in \mathbb{Z}_{[0, N-1]},\\
        & \hat{\mathbf{s}}[0] \in \bigtimes_{i=1}^{M}[\alpha_ih_\textrm{min}, (1-\beta_i)h],\\
        & \hat{\mathbf{s}}[n] \in [0, h]^M,\,\, \forall n \in \mathbb{Z}_{[1, N-1]},
    \end{aligned}
    \label{Equation:MPCProjected}
\end{equation}
to determine an approximate solution for the first control $\hat{\mathbf{s}}[0]$.

The vectors $\bm{\alpha}$ and $\bm{\beta}$ are initialized as $\mathbf{0}_M$ to fully ignore the 
infeasible set for the pulse duration, $(0, h_{\textrm{min}})$, in the first iteration of the 
algorithm. Then, at each iteration, the algorithm solves \eqref{Equation:MPCProjected} using the 
current $\bm{\alpha}$ and $\bm{\beta}$ values and checks if the first control of the solution, 
$\hat{\mathbf{s}}[0]$, is feasible for the original problem \eqref{Equation:MPC}. If it does, the 
algorithm outputs $\hat{\mathbf{s}}[0]$. Otherwise, the algorithm computes the projection of 
$\hat{\mathbf{s}}[0]$ onto the feasible set $(\{0\} \cup [h_\text{min}, h])^M$, denoted as 
$\Bar{\mathbf{s}}$. For each activation time $\hat{s}_i[0]$ that is infeasible for 
\eqref{Equation:MPC} (i.e., $\hat{s}_i[0] \in (0, h_\text{min})$), the projected component 
$\Bar{s}_i$ is either $\Bar{s}_i = 0$ or $\Bar{s}_i = h_\text{min}$. Thus, the algorithm locks 
$\hat{s}_i[0]$ to either $\{0\}$ if $\Bar{s}_i = 0$ or to the interval $[h_\text{min}, h]$ if 
$\Bar{s}_i = h_\text{min}$, by updating $\alpha_i$ and $\beta_i$ accordingly. On the other hand, 
for each feasible activation time $\hat{s}_i[0]$, we have that $\hat{s}_i[0] = \Bar{s}_i$, and the 
corresponding $\alpha_i$ and $\beta_i$ values remain unchanged. This process repeats until a 
feasible solution $\hat{\mathbf{s}}[0]$ is found, terminating after at most $M+1$ iterations, which 
would match the case in which only one pair $(\alpha_i, \beta_i)$ is updated per iteration. This 
process is further detailed in Algorithm \ref{Algorithm:Projected}.

\subsection{Relaxed Algorithm}

The second algorithm, denoted as \textbf{Relaxed}, solves \eqref{Equation:MPCProjected} just once 
with $\bm{\alpha} = \bm{\beta} = \mathbf{0}_M$, fully ignoring the infeasible set for the pulse 
duration. The first control of the solution, $\hat{\mathbf{s}}[0]$, is then projected onto the set 
$(\{0\} \cup [h_\text{min}, h])^M$ to return a feasible solution for \eqref{Equation:MPC}. 
Algorithm \ref{Algorithm:Relaxed} details the \textbf{Relaxed} solver.

\begin{algorithm}[H]
    \caption{Projected Solver}
    \label{Algorithm:Projected}
    \begin{algorithmic}[1]
        \STATE \textbf{Input}: $\mathbf{x}[k]$
        \STATE \textbf{Output}: $\hat{\mathbf{s}}[0]$
        \STATE \textbf{Set}: $\bm{\alpha} \leftarrow \mathbf{0}_M$, 
        $\bm{\beta} \leftarrow \mathbf{0}_M$
        \STATE \textbf{For} $\text{iter} \in \mathbb{Z}_{[0, M]}$:
        \STATE \quad Solve \eqref{Equation:MPCProjected} and set $\hat{\mathbf{s}}[0]$ to the 
        solution
        \STATE \quad \textbf{If} $\hat{\mathbf{s}}[0] \in (\{0\} \cup [h_\text{min}, h])^M$:
        \STATE \quad\quad \textbf{Return} $\hat{\mathbf{s}}[0]$
        \STATE \quad\textbf{EndIf}
        \STATE \quad $\underset{\mathbf{s} \in (\{0\} \cup [h_\text{min}, h])^M\hphantom{a}}
        {\Bar{\mathbf{s}} \leftarrow\,\;\text{arg min}\;\,\, 
        \|\hat{\mathbf{s}}[0]-\mathbf{s}\|}$
        \STATE \quad \textbf{For} $i \in \{j \in \mathbb{Z}_{[1, M]}: \hat{s}_j[0] \in 
        (0, h_{\text{min}})\}$:
        \STATE \quad\quad $\alpha_i \leftarrow \frac{\Bar{s}_i}{h_\text{min}}$, 
        $\beta_i \leftarrow | 1 - \alpha_i |$  \hfill\COMMENT{$\Bar{s}_i \in \{0, h_\text{min}\}$}
        \STATE \quad\textbf{EndFor}
        \STATE \textbf{EndFor}
    \end{algorithmic}
\end{algorithm}

\begin{algorithm}[H]
    \caption{Relaxed Solver}
    \label{Algorithm:Relaxed}
    \begin{algorithmic}[1]
        \STATE \textbf{Input}: $\mathbf{x}[k]$
        \STATE \textbf{Output}: $\hat{\mathbf{s}}[0]$
        \STATE Solve \eqref{Equation:MPCProjected} with $\bm{\alpha} = \bm{\beta} = \mathbf{0}_M$, 
        set $\hat{\mathbf{s}}[0]$ to the solution
        \STATE $\underset{\mathbf{s} \in (\{0\} \cup [h_\text{min}, h])^M\hphantom{a}}{\hat{\mathbf{s}}[0] \leftarrow\,\;\text{arg min}\;\,\, \|\hat{\mathbf{s}}[0]-\mathbf{s}\|}$
        \STATE \textbf{Return} $\hat{\mathbf{s}}[0]$
    \end{algorithmic}
\end{algorithm}

The \textbf{Relaxed} algorithm represents the simplest approach, considered both because it is 
computationally the least expensive and also a natural conceptual extension of the 
\textbf{Projected} algorithm. Furthermore, in \cite{malyuta2020lossless}, a comparable 
convexification technique has been shown to produce the optimal solution for a class of optimal 
control problems with similar mixed-integer constraints. However, the problem addressed in 
our paper has a different structure from the problems considered in \cite{malyuta2020lossless} and 
the actuation constraints do not verify some of the conditions that ensure the optimality of the 
\textbf{Relaxed} solution with respect to the original problem. In this paper, we introduce a 
similar convexification approach for the class of problems \eqref{Equation:MPC}, with the addition 
of the projection onto the feasible set to ensure a solution that satisfies the constraints of the 
original problem.


\section{Simulation Results} \label{Section:SimulationResults}

In this section, the efficacy of the \gls{MPC} controller is assessed for the different solvers 
through simulation examples obtained in a MATLAB environment. For simplicity, and without loss of 
generality, we assume the target's orbital plane is aligned with the $zx$-plane of the 
\gls{ECI} frame. Consequently, the motion of the target and the rotation matrix are described by
\begin{equation} 
    \begin{aligned}
    \mathbf{r}_T^E(t) = 
    \scalebox{0.8}{$
    \begin{bmatrix}
        R_T\cos(\omega t)\\
        0\\
        R_T\sin(\omega t)
    \end{bmatrix}$},\,\,
    \mathbf{R}(t) = 
    \scalebox{0.8}{$
    \begin{bmatrix}
        -\sin(\omega t) & 0 & -\cos(\omega t)\\
        0 & 1 & 0\\
        \cos(\omega t) & 0 & -\sin(\omega t)
    \end{bmatrix}$},
    \end{aligned}
\end{equation}
with $\bm{\omega} = [0\,\,0\,\,\omega]^\top$ and $\omega = \sqrt{Gm_ER_T^{-3}}$. The parameters 
used in the simulations are presented in Table \ref{Table:SimulationParameters}.

The simulation results presented in this section were obtained using the Gurobi \cite{gurobi} 
solver as the \textbf{Standard} algorithm. All computations were executed on a single desktop 
computer equipped with an AMD Ryzen 5 3600 @ \SI{3.60}{\giga\hertz} processor and 
\SI{32}{\giga\byte} of RAM. The numerical integration of the complete dynamics was conducted using 
the Runge–Kutta method through the \textit{ode45} implementation in MATLAB.

\begin{table}[H]
    \centering
    \caption{Default simulation parameters}
    \label{Table:SimulationParameters}
    \begin{threeparttable}
    \begin{tabular}{@{}lll@{}}
        \toprule
        Parameter       & Value                                 & Description                    \\ 
        \midrule
        $G$             & \SI{6.674e-11}{m^3.kg^{-1}.s^{-2}}    & Gravitational constant 
        \hphantom{Univer}  \\
        $m_E$           & \SI{5.972e24}{kg}                     & Earth's mass                   \\
        $R_T$           & \SI{7171}{km}                         & Target's orbital radius        \\
        $m_C$           & \SI{2000}{kg}                         & Chaser's mass                  \\
        $M$             & 6                                     & Number of thrusters            \\
        $\mathbf{f}_i$  & 1000$\mathbf{w}_i$ [N] \tnote{*}      & Thrust forces                  \\
        $\mathbf{Q}$    & $\text{diag}(\mathbf{1}_6)$           & State error weight matrix      \\
        $h_\text{min}$  & \SI{5}{s}                             & Minimum activation time        \\
        $h$             & \SI{10}{s}                            & Sampling period                \\
        $T_\text{sim}$  & \SI{3600}{s}                          & Simulation time                \\
        $s_0$           & $h/2$                                 & Linearization point            \\
        $\mathbf{x}(0)$ & $[0\,\,0\,\,\SI{100}{\kilo\meter}\,\,0\,\,0\,\,0]^\top$  
        & Chaser's initial state  \\
        \bottomrule
    \end{tabular}%
    \begin{tablenotes} \scriptsize
        \item [*] $\mathbf{w}_1 = -\mathbf{w}_4 = [1\,0\,0]^\top$, $\mathbf{w}_2 = -\mathbf{w}_5 = 
        [0\,1\,0]^\top$, $\mathbf{w}_3 = -\mathbf{w}_6 = [0\,0\,1]^\top$.
    \end{tablenotes}
    \end{threeparttable}
\end{table}

Assessing quantitative metrics is useful for drawing conclusions about the \gls{MPC} controller's  
performance. In this scenario, we consider two evaluation metrics:
\begin{enumerate}
    \item \textbf{Mission Time:} the mission time is the time $T_\text{mission}$ such that the 
    distance between the chaser and the target does not exceed \SI{1}{km} for all 
    $t \geq T_\text{mission}$. 
    \item \textbf{Fuel Consumption}: the fuel consumption refers to the total amount of fuel spent 
    by the chaser spacecraft. Since we assume identical thrusters, except for their direction, the 
    fuel consumed by each thruster is given by the total duration of its activation. The overall 
    fuel consumption is the sum of the individual fuel consumption of each thruster, measured in 
    seconds.
\end{enumerate}

\subsection{Variation of the Minimum Activation Time}

We begin by briefly assessing the influence of the minimum activation time on the trajectories. In 
Fig. \ref{Figure:VariationActivationTime}, we present the resulting trajectories and activation 
profiles obtained with the \textbf{Standard} algorithm for $h_\text{min} = \SI{0}{s}$, 
$h_\text{min} = \SI{2}{s}$, and $h_\text{min} = \SI{4}{s}$. Since the trajectories obtained with 
the other two algorithms are indistinguishable from those in Fig. 
\ref{Figure:VariationActivationTime}, we present only those obtained with the \textbf{Standard} 
algorithm. The trajectories are displayed in $zx$-plane of the \gls{LVLH} frame, showcasing only 
the activation profiles of the thrusters along this plane. Additionally, Table 
\ref{Table:VariationActivationTime} presents the fuel consumption, mission time, and the 
accumulated \gls{MPC} solve time for each simulation.

Examining Fig. \ref{Figure:VariationActivationTime} and Table \ref{Table:VariationActivationTime} 
reveals that the minimum activation time has a minimal impact on the overall performance of the 
controller. The central part of the maneuver, requiring certain thrusters to remain continuously on 
for extended periods, remains unaffected. Notably, for $h_\text{min} = \SI{0}{s}$, the algorithm 
addresses a linear problem without integer constraints, leading to significantly lower computation 
times. These findings suggest the feasibility of using a convex relaxation, which the 
\textbf{Projected} and \textbf{Relaxed} algorithms are designed around.
\newcommand{\xchaserinit}{
    \begin{tikzpicture}
        \protect\draw[black, thick] (-0.05,-0.05) -- (0.05,0.05);
        \protect\draw[black, thick] (-0.05,0.05) -- (0.05,-0.05);
    \end{tikzpicture}
}
\newcommand{\xtarget}{
    \begin{tikzpicture}
        \protect\draw[black, thick] (0,0) circle (2pt);
    \end{tikzpicture}
}

\vspace{-2.15mm}
\begin{figure}[H] 
    \centering 
    \subfloat[Trajectory]{
        \includegraphics[width=0.495\linewidth]{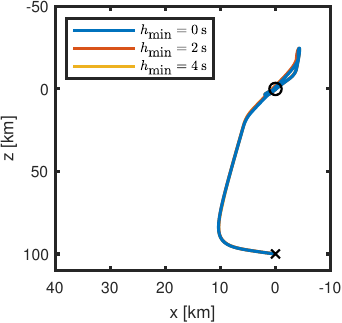}}
    \subfloat[Actuation]{
        \includegraphics[width=0.479\linewidth]{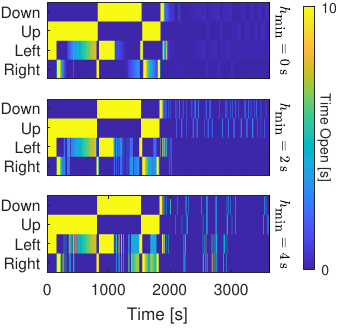}}
    \caption{Resulting trajectories and activation profiles obtained using the \textbf{Standard} 
    algorithm for different values of the minimum activation time. Chaser's initial position 
    \xchaserinit, target position \xtarget.}
    \label{Figure:VariationActivationTime}
\end{figure}

\vspace{-2.15mm}
\begin{table}[H]
\centering
\caption{Mission times and fuel consumption obtained for different values of the minimum activation 
time with $N = 10$}
\label{Table:VariationActivationTime}
\begin{tabular}{@{}llll@{}}
\toprule
Min Activation Time & Fuel Cons.& Mission Time     & Acc. Solve Time\hphantom{uni}\\\midrule
$h_\text{min} = \SI{0}{s}$      & \SI{3070.49}{s}  & \SI{1930}{s} & \SI{4.50}{s}  \\
$h_\text{min} = \SI{2}{s}$      & \SI{2931.49}{s}  & \SI{1890}{s} & \SI{27.19}{s} \\
$h_\text{min} = \SI{4}{s}$      & \SI{3068.53}{s}  & \SI{1890}{s} & \SI{21.62}{s} \\ \bottomrule
\end{tabular}
\end{table}

\subsection{Variation of the Horizon Length}

\begin{figure}[H] 
    \centering 
    \subfloat[$N = 5$ - Trajectory]{
        \includegraphics[width=0.495\linewidth]{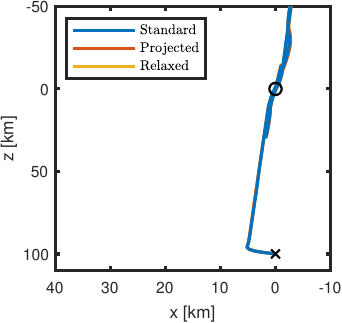}}
    \subfloat[$N = 5$ - Actuation]{
        \includegraphics[width=0.479\linewidth]{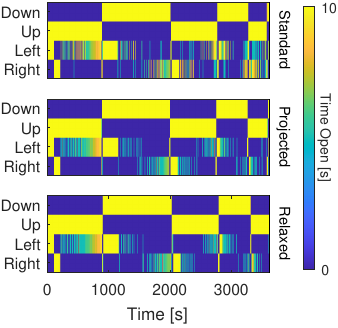}}\\
    \subfloat[$N = 10$ - Trajectory]{
        \includegraphics[width=0.495\linewidth]{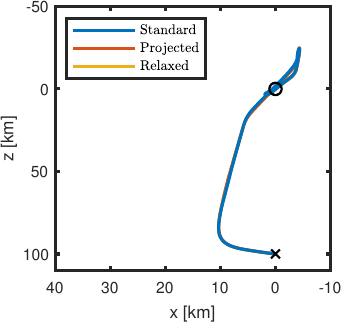}}
    \subfloat[$N = 10$ - Actuation]{
        \includegraphics[width=0.479\linewidth]{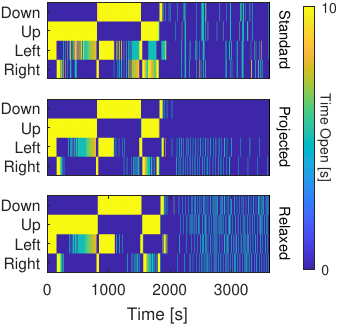}}\\
    \subfloat[$N = 15$ - Trajectory]{
        \includegraphics[width=0.495\linewidth]{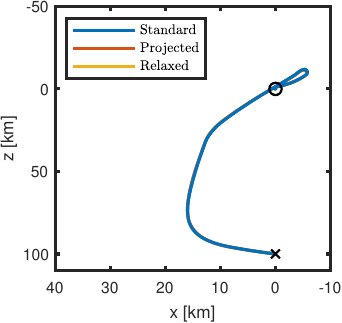}}
    \subfloat[$N = 15$ - Actuation]{
        \includegraphics[width=0.479\linewidth]
        {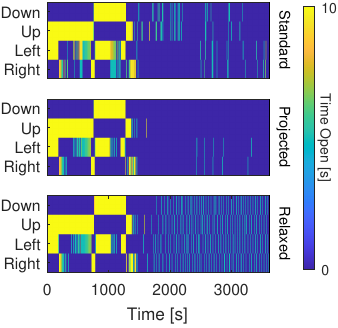}}\\
    \subfloat[$N = 100$ - Trajectory]{
        \includegraphics[width=0.495\linewidth]{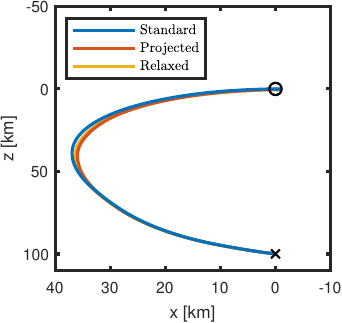}}
    \subfloat[$N = 100$ - Actuation]{
        \includegraphics[width=0.479\linewidth]{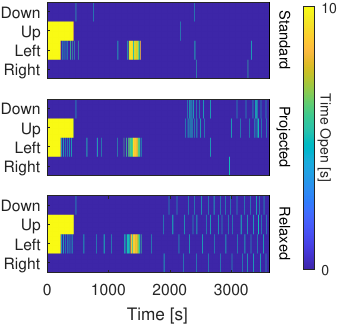}}
    \caption{Resulting trajectories and activation profiles obtained using each algorithm 
    for different horizon lengths. Chaser's initial position \xchaserinit, target position 
    \xtarget.}
    \label{Figure:ResultsTrajectories}
\end{figure}

In order to evaluate the influence of the prediction horizon on the algorithms, the three solver 
algorithms were applied to the rendezvous scenario employing different horizon lengths. Fig. 
\ref{Figure:ResultsTrajectories} displays the resulting trajectories and activation profiles for 
$N = 5$, $N = 10$, $N = 15$, and $N = 100$. Additionally, Table \ref{Table:MissionTimeFuelSpent} 
presents the fuel consumption, mission time, and the accumulated \gls{MPC} solve time for each 
simulation.

\vspace{-2.4mm}

\begin{table}[H]
    \centering
    \caption{Mission times and fuel consumption obtained using each algorithm for different 
    horizon lengths}
    \label{Table:MissionTimeFuelSpent}
    \begin{tabular}{@{}lllll@{}}
    \toprule
    Horizon                 & Algorithm & Fuel Cons. & Mission Time & Acc. Solve Time \\ 
                            \midrule
    \multirow{3}{*}{$N = 5$}    & Standard & \SI{5667.33}{s} & \SI{3580}{s} & \SI{4.25}{s}  \\
                            & Projected & \SI{5186.60}{s} & \SI{3570}{s} & \SI{3.36}{s} \\
                            & Relaxed & \SI{5229.32}{s} & \SI{3590}{s} & \SI{2.01}{s}   \\
                            \midrule
    \multirow{3}{*}{$N = 10$}   & Standard & \SI{3286.42}{s} & \SI{1860}{s} & \SI{14.39}{s} \\
                            & Projected & \SI{2925.65}{s} & \SI{1890}{s} & \SI{3.84}{s} \\
                            & Relaxed & \SI{2885.57}{s} & \SI{1880}{s} & \SI{2.89}{s}   \\
                            \midrule
    \multirow{3}{*}{$N = 15$}   & Standard & \SI{2470.85}{s} & \SI{1420}{s} & \SI{27.86}{s} \\
                            & Projected & \SI{2252.42}{s} & \SI{1420}{s} & \SI{4.64}{s} \\
                            & Relaxed & \SI{2299.30}{s} & \SI{1430}{s} & \SI{3.10}{s}   \\
                            \midrule
    \multirow{3}{*}{$N = 100$}  & Standard & \SI{777.13}{s} & \SI{1400}{s} & \SI{3791.84}{s} \\
                            & Projected & \SI{791.25}{s} & \SI{1430}{s} & \SI{30.69}{s} \\
                            & Relaxed & \SI{808.94}{s} & \SI{1430}{s} & \SI{23.08}{s} \\
                            \bottomrule
    \end{tabular}
\end{table}

The results suggest that exploring the mixed-integer solution space does not yield substantial 
benefits in such scenarios, as both the \textbf{Projected} and \textbf{Relaxed} algorithms 
produce trajectories that are very similar to the optimal solution. In addition, Table
\ref{Table:MissionTimeFuelSpent} reinforces this observation, as the fuel consumption and mission 
time across different values of $N$ are very similar for all three algorithms. Remarkably, the 
\textbf{Projected} algorithm, which partially explores the mixed-integer solution space, fails to 
generate significantly better trajectories than the \textbf{Relaxed} algorithm, which does not 
explore the mixed-integer solution space at all. With a sufficiently large prediction horizon, such 
as $N=100$, the results resemble a Hohmann transfer orbit - the (two impulse) maneuver 
using the lowest possible amount of energy to transfer a spacecraft between orbits.

The solutions produced by these algorithms closely resemble the optimal solution, as observed in 
the activation profiles and consequently in the trajectories themselves. The substantial 
accumulated computation time of the \textbf{Standard} algorithm makes it impractical not only for 
real-time applications but also for simulations employing extended prediction horizons. Conversely, 
the \textbf{Projected} and \textbf{Relaxed} algorithms are much more computationally efficient, 
rendering them a more suitable option for real-time applications.

\subsection{Computation Times}

\begin{figure}[tp] 
    \centering 
    \subfloat[$N = 5$ - Histogram]{
        \includegraphics[width=0.47\linewidth]{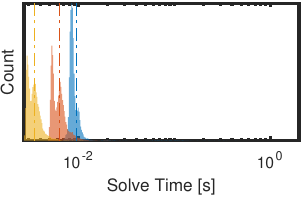}}
        \hfill
    \subfloat[$N = 5$ - Profile]{
        \includegraphics[width=0.5\linewidth]{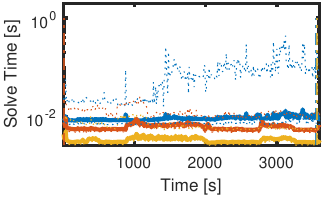}}\\
        [-0.2cm]
    \subfloat[$N = 10$ - Histogram]{
        \includegraphics[width=0.47\linewidth]{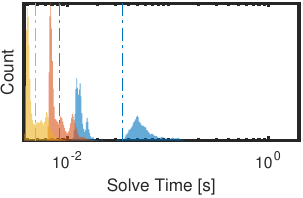}}
        \hfill
    \subfloat[$N = 10$ - Profile]{
        \includegraphics[width=0.5\linewidth]{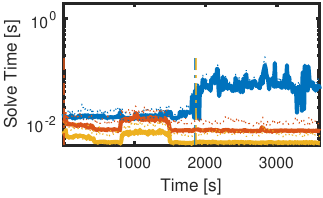}}\\
        [-0.2cm]
    \subfloat[$N = 15$ - Histogram]{
        \includegraphics[width=0.47\linewidth]{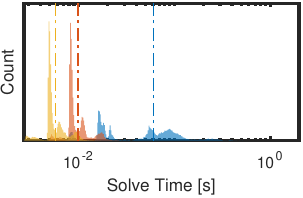}}
        \hfill
    \subfloat[$N = 15$ - Profile]{
        \includegraphics[width=0.5\linewidth]{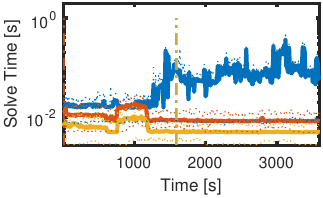}}
    \caption{Computation times obtained using each algorithm for different horizons -
    \textbf{Standard} (blue), \textbf{Projected} (red), \textbf{Relaxed} (yellow). Mean 
    solve time (dash-dotted, histograms), mission time (dash-dotted, profiles), maximum solve 
    time (dotted, profiles).} 
    \label{Figure:ResultsTimes}
\end{figure}

For a deeper analysis of the computational cost of each algorithm, we conducted 100 simulations 
using each algorithm for $N = 5$, $N = 10$, and $N = 15$. Fig. \ref{Figure:ResultsTimes} depicts 
the resulting histograms and the mean solve times over time. Additionally, Table 
\ref{Table:SolverTimes} provides the values of the mean computation times, as well as the 
\SI{95}{\percent} and \SI{99}{\percent} percentiles of these times.

As presented in Fig. \ref{Figure:ResultsTimes} and Table \ref{Table:SolverTimes}, the 
\textbf{Standard} algorithm exhibits a significantly greater average computation time than the 
other two algorithms across all values of $N$. The computation times of the \textbf{Standard} 
algorithm also exhibit a greater dispersion, as demonstrated by their wider distributions. On the 
other hand, the \textbf{Projected} and \textbf{Relaxed} algorithms differ by less than an order of 
magnitude in terms of average computation time, with the \textbf{Projected} algorithm taking 
slightly longer than the \textbf{Relaxed} algorithm. This discrepancy is anticipated, as 
the \textbf{Projected} algorithm solves multiple optimization problems per \gls{MPC} iteration, 
while the \textbf{Relaxed} algorithm solves only one per iteration. The \textbf{Projected} and 
\textbf{Relaxed} algorithms also display similar and narrower distributions of the solve times.

The distributions shown in Fig. \ref{Figure:ResultsTimes} exhibit a bimodal pattern, which 
becomes more pronounced as $N$ increases, particularly for the \textbf{Standard} algorithm. This 
behavior can be understood by observing the temporal profiles of the mean solver times in Fig. 
\ref{Figure:ResultsTimes}. One can observe that the solve times for each algorithm are not constant 
throughout the simulation; instead, they vary significantly, especially for the \textbf{Standard} 
algorithm. Interestingly, as the simulation progresses and the chaser approaches the target, the 
solve times for the \textbf{Standard} algorithm increase. In contrast, the solve times for the 
other two algorithms exhibit less variation throughout the simulation and, in fact, tend to 
decrease as the simulation advances.

\begin{table}[H]
\centering
\caption{Computation times obtained using each algorithm \\for different horizon lengths}
\label{Table:SolverTimes}
\begin{tabular}{@{}lllll@{}}
\toprule
Horizon & Algorithm & Mean & 95\% Percentile & 99\% Percentile \hphantom{Fil}\\ \midrule
\multirow{3}{*}{$N = 5$} & Standard & \SI{9.46}{\milli\second} & \SI{10.67}{\milli\second} & 
\SI{17.79}{\milli\second}\\
    & Projected & \SI{6.27}{\milli\second} & \SI{8.59}{\milli\second} & \SI{10.79}{\milli\second}\\
    & Relaxed & \SI{3.44}{\milli\second} & \SI{4.53}{\milli\second} & \SI{5.74}{\milli\second}\\ 
    \midrule
\multirow{3}{*}{$N = 10$} & Standard & \SI{35.07}{\milli\second} & \SI{84.22}{\milli\second} & 
\SI{124.13}{\milli\second}\\
    & Projected & \SI{8.27}{\milli\second} & \SI{12.30}{\milli\second} & \SI{14.88}{\milli\second}\\
    & Relaxed & \SI{4.76}{\milli\second} & \SI{6.75}{\milli\second} & \SI{7.80}{\milli\second}\\ 
    \midrule
\multirow{3}{*}{$N = 15$}& Standard &\SI{60.80}{\milli\second} & \SI{163.96}{\milli\second} & 
\SI{280.52}{\milli\second}\\
    & Projected &\SI{9.91}{\milli\second} & \SI{17.31}{\milli\second} & \SI{19.55}{\milli\second}\\
    & Relaxed &\SI{5.79}{\milli\second} & \SI{9.58}{\milli\second} & \SI{10.91}{\milli\second}\\ 
    \bottomrule
\end{tabular}
\end{table}


\section{Conclusion \& Future Research} \label{Section:Conclusion}

This paper focuses on the application of \gls{MPC} to spacecraft rendezvous, addressing the 
computational challenges arising from mixed-integer constraints on the actuation. We propose 
two solver algorithms that efficiently approximate the optimal solution in significantly less time 
than standard \gls{MIP} solvers, rendering them suitable for real-time applications. Extensions may 
include addressing elliptical orbits and safety concerns.


\bibliographystyle{ieeetr}
\bibliography{references}

\begin{thebibliography}{10}

\bibitem{bonnal2013active}
C.~Bonnal, J.-M. Ruault, and M.-C. Desjean, ``Active debris removal: {{Recent}} progress and current trends,'' {\em Acta Astronautica}, vol.~85, pp.~51--60, Apr. 2013.

\bibitem{arney2021orbit}
D.~Arney, R.~Sutherland, J.~Mulvaney, D.~Steinkoenig, C.~Stockdale, and M.~Farley, ``On-orbit {{Servicing}}, {{Assembly}}, and {{Manufacturing}} ({{OSAM}}) {{State}} of {{Play}}, 2021 {{Edition}},'' White Paper 20210022660, {OSAM National Initiative}, Oct. 2021.

\bibitem{fehse2003automated}
W.~Fehse, {\em Automated {{Rendezvous}} and {{Docking}} of {{Spacecraft}}}.
\newblock {Cambridge University Press}, Nov. 2003.

\bibitem{bezerra2021optimal}
J.~A. Bezerra and D.~A. Santos, ``Optimal exact control allocation for under-actuated multirotor aerial vehicles,'' {\em IEEE Control Systems Letters}, vol.~6, pp.~1448--1453, 2021.

\bibitem{ankersen2005optimization}
F.~Ankersen, S.-F. Wu, A.~Aleshin, A.~Vankov, and V.~Volochinov, ``Optimization of spacecraft thruster management function,'' {\em Journal of guidance, control, and dynamics}, vol.~28, no.~6, pp.~1283--1290, 2005.

\bibitem{hartley2015tutorial}
E.~N. Hartley, ``A tutorial on model predictive control for spacecraft rendezvous,'' in {\em 2015 {{European Control Conference}} ({{ECC}})}, pp.~1355--1361, July 2015.

\bibitem{weiss2015model}
A.~Weiss, M.~Baldwin, R.~S. Erwin, and I.~Kolmanovsky, ``Model predictive control for spacecraft rendezvous and docking: Strategies for handling constraints and case studies,'' {\em IEEE Transactions on Control Systems Technology}, vol.~23, pp.~1638--1647, July 2015.

\bibitem{silvestre2023model}
D.~Silvestre and G.~Ramos, ``Model {{Predictive Control With Collision Avoidance}} for {{Unknown Environment}},'' {\em IEEE Control Systems Letters}, vol.~7, pp.~2821--2826, 2023.

\bibitem{zhu2018robust}
S.~Zhu, R.~Sun, J.~Wang, J.~Wang, and X.~Shao, ``Robust model predictive control for multi-step short range spacecraft rendezvous,'' {\em Advances in Space Research}, vol.~62, no.~1, pp.~111--126, 2018.

\bibitem{leomanni2021sum}
M.~Leomanni, G.~Bianchini, A.~Garulli, and R.~Quartullo, ``Sum-of-{Norms} {{Periodic Model Predictive Control}} for {{Space Rendezvous}},'' {\em IEEE Transactions on Control Systems Technology}, vol.~30, no.~3, pp.~1311--1318, 2021.

\bibitem{brizguidance}
J.~Briz, N.~Paulino, P.~Louren{\c c}o, P.~Cachim, E.~{Forgues-Mayet}, L.~Ferreira, A.~Groth, E.~Papadopoulos, G.~Rekleitis, K.~Nanos, and V.~Preda, ``Guidance, {{Navigation}}, and {{Control}} of {{In-Orbit Assembly}} of {{Large Antennas}} {\textendash} technologies and approach for {{IOANT}},'' in {\em Proceedings of the 41st {{ESA Antenna Workshop}} on {{Large Deployable Antennas}}}, ({Noordwijk, The Netherlands}), {ESA}, Sept. 2023.

\bibitem{lovelly:2018:BenchmarkingAnalysisSpaceGrade}
T.~M. Lovelly, T.~W. Wise, S.~H. Holtzman, and A.~D. George, ``Benchmarking {{Analysis}} of {{Space-Grade Central Processing Units}} and {{Field-Programmable Gate Arrays}},'' {\em Journal of Aerospace Information Systems}, vol.~15, pp.~518--529, Aug. 2018.

\bibitem{accikmecse2011lossless}
B.~A{\c{c}}{\i}kme{\c{s}}e and L.~Blackmore, ``Lossless convexification of a class of optimal control problems with non-convex control constraints,'' {\em Automatica}, vol.~47, no.~2, pp.~341--347, 2011.

\bibitem{malyuta2020lossless}
D.~Malyuta and B.~A{\c{c}}ikme{\c{s}}e, ``Lossless convexification of optimal control problems with semi-continuous inputs,'' {\em IFAC-PapersOnLine}, vol.~53, no.~2, pp.~6843--6850, 2020.

\bibitem{hespanhol2019structure}
P.~Hespanhol, R.~Quirynen, and S.~Di~Cairano, ``A structure exploiting branch-and-bound algorithm for mixed-integer model predictive control,'' in {\em 2019 18th European Control Conference (ECC)}, pp.~2763--2768, IEEE, 2019.

\bibitem{gurobi}
{Gurobi Optimization, LLC}, ``{Gurobi Optimizer Reference Manual},'' 2022.

\bibitem{bemporad2018numerically}
A.~Bemporad and V.~V. Naik, ``A numerically robust mixed-integer quadratic programming solver for embedded hybrid model predictive control,'' {\em IFAC-PapersOnLine}, vol.~51, no.~20, pp.~412--417, 2018.

\bibitem{Yang2018}
Y.~Yang, ``Coupled orbital and attitude control in spacecraft rendezvous and soft docking,'' {\em Proceedings of the Institution of Mechanical Engineers, Part G: Journal of Aerospace Engineering}, vol.~233, p.~3109–3119, Aug. 2018.

\bibitem{brown:1996:SpacecraftPropulsion}
C.~D. Brown, {\em Spacecraft {{Propulsion}}}.
\newblock {Washington DC}: {American Institute of Aeronautics and Astronautics}, Jan. 1996.

\end{thebibliography}

\end{document}